# Collaborative Feature Maps of Networks and Hosts for AI-driven Intrusion Detection

Jinxin Liu, Murat Simsek, Burak Kantarci
School of Electrical Engineering and Computer Science
University of Ottawa
Ottawa, Canada
{jliu367, murat.simsek, burak.kantarci} @uottawa.ca

Mehran Bagheri, Petar Djukic
AI and Analytics
Ciena Corp.
Ottawa, Canada
{mbagheri, pdjukic} @ciena.com

*Abstract*—Intrusion Detection Systems (IDS) are critical security mechanisms that protect against a wide variety of network threats and malicious behaviors on networks or hosts. As both Network-based IDS (NIDS) or Host-based IDS (HIDS) have been widely investigated, this paper aims to present a Combined Intrusion Detection System (CIDS) that integrates network and host data in order to improve IDS performance. Due to the scarcity of datasets that include both network packet and host data, we present a novel CIDS dataset formation framework that can handle log files from a variety of operating systems and align log entities with network flows. A new CIDS dataset named SCVIC-CIDS-2021 is derived from the meta-data from the well-known benchmark dataset, CIC-IDS-2018 by utilizing the proposed framework. Furthermore, a transformer-based deep learning model named CIDS-Net is proposed that can take network flow and host features as inputs and outperform baseline models that rely on network flow features only. Experimental results to evaluate the proposed CIDS-Net under the SCVIC-CIDS-2021 dataset support the hypothesis for the benefits of combining host and flow features as the proposed CIDS-Net can improve the macro F1 score of baseline solutions by 6.36% (up to 99.89%).

## I. Introduction

Intrusion Detection Systems (IDS) are instrumental for the protection of network and credential assets [1]. IDS can be broadly classified into two types based on their analysis behavior: NIDS and HIDS, the usefulness of which has been demonstrated via extensive research [2]. A NIDS is typically deployed at tactical points on a network to monitor network traffic from all network devices and to identify malicious activities and policy violations [3]. NIDS's decision-making engine can be rule/policy-based or machine/deep learning-based which is capable of mining patterns from massive amounts of network traffic to make precise decisions. On the other hand, a HIDS is deployed on servers and end devices. HIDS can be categorized as system log-based, system call-based, Windows registry-based, or file system-based, depending on the data source [4]. Both NIDS and HIDS have undergone substantial investigation, and a wide variety of models have been presented to reduce the model's bias and variance in order to increase the performance. Nonetheless, decomposing machine learning errors reveal that even the optimal machine learning models cannot get around data noise, and that every model will eventually approach its theoretical limit due to data noise [5]. Thus, we hypothesize that combining features and information from multiple domains is a viable approach for addressing this fundamental problem.

As illustrated in [3], few researchers combine network and host-based data, and these studies manually extract a restricted number of host-based features, such as root login and shell count. However, because the majority of the host-based information is unstructured/text-based, a large amount of data is not properly utilized during this process. As a result, processing such a large amount of unstructured data and aligning host and network-based data are extremely difficult. Thus, this study incorporates Natural Language Processing (NLP) into Combined IDS (CIDS) for the first time and makes extensive use of network and host-based data to considerably increase IDS detection performance. To do so, the paper begins with the proposal of a novel framework to create CIDS datasets that are capable of processing system logs from a variety of different Operating Systems (OS) and aligning log entries with network flows. Additionally, a flexible transformer-based deep learning model is proposed to incorporate network flow and system log data of varying shapes and dimensions with the objective of making better predictions in comparison to a model that only uses network flow features.

This work proposes CIDS containing a dataset formation framework and a deep learning based intrusion detection model named CIDS-Net which combines network flow and host based features. The main contributions of this paper are five-fold: 1) To the best of our knowledge, this is the first attempt in the CIDS field to use a deep learning model to directly extract features from host-based data rather than manually mining arbitrary features. 2) A novel framework for creating CIDS datasets is proposed that considers various host-based entities under various Operating Systems (OS) with network flow features 3) A CIDS dataset named SCVIC-CIDS-2021 is created using the meta-data from the well-known benchmark dataset CIC-IDS-2018 [6] (e.g., PCAP and log files). 4) This paper offers CIDS-Net capable of including network flow and host-based features of different shapes and dimensions. 5) A new loss function is proposed, named *C Loss*, to assist CIDS in employing more complicated models such as

TabNet and training CIDS and NIDS model simultaneously.

The rest of the paper is structured as follows. Section II provides an overview of NIDS and HIDS. Section III describes the methodology which includes the framework for creating CIDS datasets and the CIDS-Net. Section IV presents experimental results and discussions around them. Finally, in Section V, conclusions are drawn to regarding the contributions of this study.

## II. RELATED WORK

This section summarizes recent works involving the combination of network and host-based data.

Several IDS benchmarks incorporating network- and host-based data are created and updated following the publication of DARPA98/99 to include more recent attacks and more realistic network topologies and traffic flows. Previous works of relevance are arranged chronologically as follows:

KDD98/99 [7] and NSL-KDD [8] provide both network and host-based features, but have been criticized for their unrealistic traffic and out-of-date attack; also, only limited features are taken from the meta-data. Kent et al. [9] collect real-time traffic and logs from the internal computer network of Los Alamos National Laboratory. Although network and host-based features are retrieved, they are not aligned, and the number of features is limited; for example, only three network- and five host-based features are available. Turcotte et al. [10] collect network traffic and Windows events from the enterprise network of Los Alamos National Laboratory over a 90-day period and create the Unified Host and Network Dataset. The dataset represents real-world network distributions; however, it is difficult to analyze and evaluate due to the lack of labels; for example, Beazley et al. [11] explore the dataset but do not identify the intrusions. Sharafaldin et al. [6] create CIC-IDS-2018, which contains 50 attackers, 420 victim machines, and 30 servers. According to McAfee's study, the majority of attacks are injected into target networks. The collection includes meta data such as PCAP files, log files, and labeling information. CICFlowMeter [6] is used to extract features from network packets but does not analyze host-based data. Vinayakumar et al. [12] try to propose a framework for integrating NIDS and HIDS, whereas the experiment part evaluates the performance of DNN on NIDS and HIDS datasets separately. Lu et al. [13] set up a simulated environment, inject four different types of network attacks, and collect network traffic and multiple types of log files from IDS devices; nonetheless, the majority of features retrieved from logs are network-related, such as source, destination, and bytes sent/received.

The studies listed above which combine network and host-based data appear to have forced host-based data into a tabular format in order to accommodate network-based features, resulting in the loss of a large amount of valuable data. Nonetheless, models from both ends (NIDS and HIDS) have improved significantly, and numerous advanced methods/models have been proposed to be more in tune with the nature of data; for instance, Du et al. [14] offer DeepLog, which makes use of a Long Short Term Memory (LSTM) network to represent log entries as time sequences and to discover anomalies in log files.

Thus, constraining them to the same shape or dimension inhibits the integration of outstanding methodologies from two fields. This work identifies the gap for a framework to create CIDS datasets with data samples of varying forms and shapes, as well as the transformer-based CIDS-Net that can use them as inputs to detect intrusion types.

## III. COMBINED INTRUSION DETECTION SYSTEM (CIDS) METHODOLOGY

The CIDS collects data from both networks and hosts. The techniques for developing and preparing the CIDS dataset are detailed in Section III-B. This work also offers the CIDS-Net to classify network intrusions with the use of host-based features in Section III-C.

### A. Problem Definition

An Intrusion Detection System (IDS) can be broadly classified into three categories: Network-based IDS (NIDS), Host-based IDS (HIDS), and Combined IDS (CIDS). An NIDS sniffs network packets generating flows, from which network-based properties are collected and input into machine/deep learning models, whereas a HIDS monitors end device activities such as system/application logs, network logs, and operating system audit trails. CIDS is considered to bridge the strengths of NIDS and HIDS. Thus, a CIDS takes its input from a dataset that contains network- and host-based features. Due to the fact that the CIDS dataset incorporates data derived from distinct fields (network and host), a novel CIDS-Net is proposed to forecast intrusions using the combined features. The following subsections discuss the CIDS dataset formation framework and model in further detail.

### B. CIDS Dataset Formation Framework

A CIDS dataset includes both network- and host-based features. This subsection presents a novel framework, named CIDS Dataset Formation Framework (CDFF), for the generation of CIDS instances. As illustrated in Fig 1, the CDFF extracts network-related features from network packets and host-related data from parsed logs from various operating systems. Due to the fact that host-based and network-based features have separate timestamps and duration characteristics, this subsection also presents an alignment algorithm to generate CIDS samples.

Network-related features are retrieved from network flows, which are a collection of packets that share the same session key (i.e., source IP, source port, destination IP, destination port, protocol, and time window). It is worth noting that each network-based sample is monitored over a period of time rather than a single packet. Each network-based feature is a statistic value that describes three aspects: the direction of a flow, feature type, and statistical characteristics.

Operating Systems (OSs) use a variety of different log formats to store host-based data; for example, Windows OS

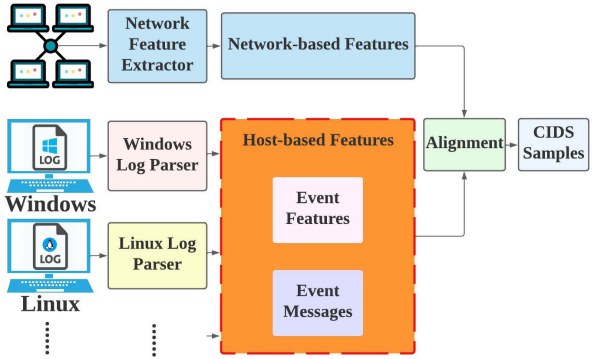

Fig. 1. CIDS Sample Generation Framework

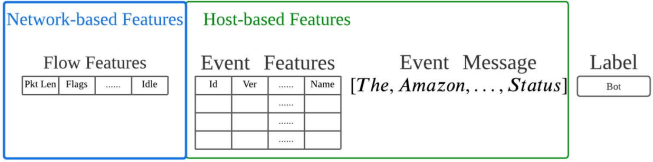

Fig. 2. The structure of a CIDS sample with four components: network-based features, event features, event messages, and label; event features and event messages are host-based features

**Algorithm 1:** Alignment of network-based ($s_n$) and host-based samples ($s_h$) to obtain combined samples ($s_c$). $S$ is a set of samples; $d$ is the time window of a network sample; $e_f$ and $e_m$ denote event feature and event message; $E_f, E_m$ are sets of $e_f$ and $e_m$

**Input:** $S_n$; $S_h$
**Output:** CIDS Samples
1  $sort(samples_h)$;
2  $S_c \leftarrow \emptyset$;
3  **foreach** $s_n \in S_n$ **do**
4   $\quad ip, d \leftarrow get\_ip\_timewindow(s_n)$;
5   $\quad E_f, E_m \leftarrow \emptyset, \emptyset$;
6   $\quad$ **foreach** $s_h \in S_h$ **do**
7   $\quad\quad$ **if** $ip \in s_h$ and $s_h.ts \in d$ **then**
8   $\quad\quad\quad e_f, e_m \leftarrow s_h$;
9   $\quad\quad\quad E_f \leftarrow \begin{bmatrix} E_f \\ e_f \end{bmatrix}$;
10  $\quad\quad\quad E_m \leftarrow E_m \frown e_m$;
11  $\quad\quad$ **end**
12  $\quad$ **end**
13  $\quad S_c \leftarrow S_c \cup (s_n, E_f, E_m)$;
14 **end**

Operating Systems (OSs) use a variety of different log formats to store host-based data; for example, Windows OS logs are saved in the EVTX format, whereas Linux OS logs are recorded in text format. While many log parsers are optimized for these formats, they can always output both structured and unstructured data (text-based event messages) from various OS logs. Structured features extracted by system event log parsers which can be directly fed into machine/deep learning models are referred to as event features (e.g., ID, version, and provider's name). However, because event messages are in plain text format and are considered unstructured data, they are hard to be analyzed by a system event log parser.

As stated previously, each network flow is monitored within a time window, but each system event is generated with a precise timestamp. Thus, each network-based sample corresponds to multiple host-based samples. Algorithm 1 is proposed to align network-based and host-based instances so that they are generated in the same time frame. In a nutshell, three important tasks are followed. Algorithm 1 seeks the host-based samples that have the same IP address and are inside the time window of the network-based instance for each network-based sample. The host-based samples' event features and event messages are then concatenated, accordingly. Each event feature instance is a vector, the output of event feature concatenation is a matrix, and event messages are connected together directly to make a longer string. Finally, each combined sample can be considered as a tuple that includes a network-based sample, "events features", and "events messages".

Fig 2 presents the final structure of a CIDS sample, which consists of four components: network-based features, events features, events messages, and a label. The "network features" is a vector of statistics observation of a flow, such as packet length, TCP flags, and packet inter-arrival Time. The "event features" is a $\mathbb{R}^{n*m}$ matrix, where $n$ is the number of event entries within time window, and $m$ is the number of event features. Since "events messages" is a string, Natural Language Processing (NLP) techniques should be utilized, which are further discussed in Section III-C

### C. CIDS-Net

Given that the CIDS dataset contains three distinct components (network-based features, event features, and event messages) with varying shapes, dimensions, and types, this work proposes a CIDS-Net to combine them and predict labels. As seen in Fig. 3, the CIDS-Net consists of three encoders (i.e., Network Feature Encoder (blue), Event Feature Encoder (yellow), and Event Message Encoder (violet)) optimized for the three separate inputs, as well as an aggregator that accepts the encoder outputs as input and predicts intrusions. Because network features are essentially tabular data, a Fully Connected Network (FCN) can be used as a network feature encoder. To improve network feature representation, deep learning model alternatives that are designed for tabular datasets can also be employed as network feature encoders. Consequently, TabNet is chosen as a viable model in this work, as it is a highly explainable model with high performance. Specifically, the network feature encoder can also be an identical layer, which means that the aggregator takes network features directly as input.

Since the essence of events features is multivariant time series, this work uses the transformer encoder as the event feature encoder, which is a state-of-the-art deep learning model for time series and NLP. Event messages, on the other hand, are readable text data that describe the specifics of event entities. The event messages in our model are initially transformed into vectors via BERT word embedding. This work employs a pre-trained BERT network fixing its weights without fine tuning to avoid potential overfitting due to plugging such a big network into the CIDS-Net. Following word embedding, event messages are transformed into a matrix of the format $\mathbb{R}^{i*j}$, where $i$ is the number of tokens in the sentences and $j$ denotes the dimension of the embedding vectors. Similar to the event feature, the event message encoder is also a transformer encoder.

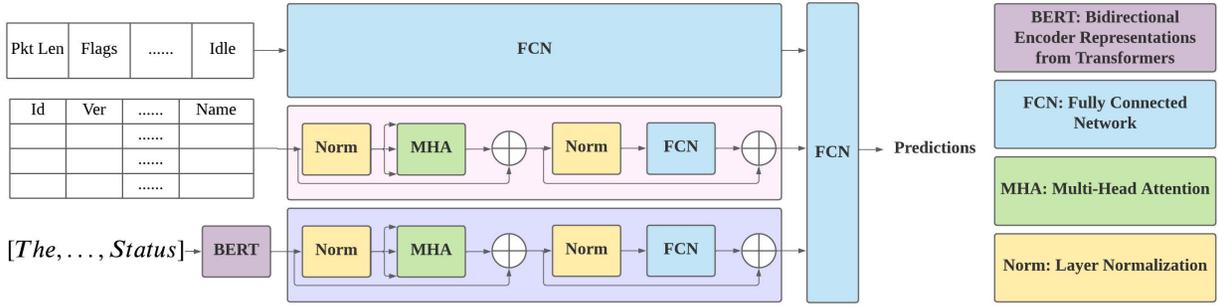

Fig. 3. The architecture of CIDS-Net

To obtain predictions, the encoded hidden states are aggregated by a FCN which is adaptable and does not have requirements for upper encoders. For FCN, a basic cross entropy loss ($CE$) function can be used, whereas this study proposes using Combined Loss (C Loss) to jointly train the network feature encoder and the whole CIDS-Net so that both of them can make predictions. y doing so, the proposed method ensures flexibility in the absence of host-based data and to incorporate more complex models. Equation 1 formulates the C Loss, where $y$ is the true label of a sample, $\hat{y}$ is the aggregator's prediction, $\hat{y}_n$ is the network feature encoder's prediction and $\alpha$ is a hyperparameter. It is worth noting that even though the FCN aggregator has no limits, the C Loss requires that the number of hidden states of the network feature encoder's output is equal to the number of classes.

$$C\_Loss = \alpha * CE(y, \hat{y}_n) + (1 - \alpha) * CE(y, \hat{y}) \quad (1)$$

Additionally, with C Loss, network feature encoders can produce complicated outputs; for example, if TabNet is plugged into the CIDS-Net as a network feature encoder, the loss of the CIDS-Net can be formulated as: $C\_loss = \alpha * (CE(\hat{y}_n, y) - \lambda * L_{sparse}) + (1 - \alpha) * CE(\hat{y}, y)$, where $L_{sparse}$ is TabNet sparsity regularization, and $\lambda$ represents the hyperparameters.

## IV. EXPERIMENTS

This section introduces the SCVIC-CIDS-2021 dataset generated by the CDFF, and presents comprehensive results and analyses of CIDS-Nets.

### A. SCVIC-CIDS-2021 Dataset

The CIDS dataset production strategy strives to extract novel features and deliver a dataset in a novel format. As network- and host-based data, realistic network settings, and recent attacks are desired, CIC-IDS-2018 dataset is utilized to form SCVIC-CIDS-2021, which contains network traffic (in PCAP format), host-based data (i.e., system logs), and labeling metadata [6]. CIC-IDS-2018 extracts only the network based features whereas the host related data are not yet processed. CIC-IDS-2018 is used as a basis and fed into our CDFF to create a new CIDS dataset named SCVIC-CIDS-2021.

Because the provided network flow-based tabular dataset in CIC-IDS-2018 lacks IP addresses and timestamps, it is regenerated using CICFlowMeter with the same settings (i.e., two minutes timewindow) and labelled according to the attack traces listed in [15]. The OS logs of CIC-IDS-2018 are acquired from Windows and Linux operating systems. The 'Get-WinEvent' cmdlet converts Windows OS logs to tabular format, and the 'ProviderName' and event messages are extracted from Linux OS logs. Algorithm 1 is used to align network- and host-based data. The CIC-IDS-2018 dataset contains more than 1 TB of network intrusion data after decompression, thus, as reported in [16], the majority of the literature that utilizes CIC-IDS-2018 only uses a subset of the dataset, such as [17], [18]. SCVIC-CIDS-2021 becomes much larger when host-based features are added. The benign traffic is randomly undersampled to 50%, as they typically have greater than 99% F1 score [19]. To prevent adding an excessive amount of missing data, SCVIC-CIDS-2021 excludes samples that lack host-based data. To overcome the concerns about this procedure, we additionally demonstrate the results when the CIDS-Net is not given host-based features. After eliminating samples that do not have host-based features, several minority classes become even rare, containing fewer than twenty samples (e.g., 'Brute Force - Web' (15), 'Brute Force - XSS' (6), 'Infiltration' (8), and 'SQL Injection' (8)). Since the limited samples will occur in the testing process, rare cases will significantly affect the deep learning model performance; hence, those rare classes are eliminated [20].

Network features are trivial as the ones in the original CIC-IDS-2018. The largest number of system event entity is 28; thus all event features are zero-padded with the shape of $\mathbb{R}^{28*8}$.

TABLE I
SCVIC-CIDS-2021 CLASS DISTRIBUTION

|  | Training Set | Test Set | Total |
|---|---|---|---|
| Benign | 308375 | 152172 | 460547 |
| Bot | 60767 | 29693 | 90460 |
| DDOS-HOIC | 137147 | 67449 | 204596 |
| DDOS-LOIC-HTTP | 39019 | 19166 | 58185 |
| DDOS-LOIC-UDP | 760 | 342 | 1102 |
| DoS-GoldenEye | 2271 | 1163 | 3434 |
| DoS-Hulk | 13388 | 6553 | 19941 |
| DoS-SlowHTTPTest | 10579 | 5351 | 15930 |
| DoS-Slowloris | 1394 | 702 | 2096 |
| FTP-BruteForce | 32222 | 15918 | 48140 |
| SSH-Bruteforce | 11143 | 5418 | 16561 |
| Sum | 617065 | 303927 | 920992 |

TABLE II
PERFORMANCE OF MACHINE LEARNING ALGORITHMS UNDER
SCVIC-CIDS-2021 WHEN ONLY NETWORK-BASED FEATURES ARE USED
(BASELINE RESULTS); LR: LOGISTIC REGRESSION; AB: ADABOOST;
NB: NAIVE BAYES; GB: GRADIENTBOOST; DT: DECISION TREE; RF:
RANDOM FOREST

|  | LR | AB | NB | TabNet | GB | DT | XGB | RF |
|---|---|---|---|---|---|---|---|---|
| Benign | 0.780 | 0.891 | 0.698 | 0.999 | 0.999 | 0.999 | 0.999 | 1.000 |
| Bot | 0.009 | 0.000 | 1.000 | 1.000 | 1.000 | 1.000 | 1.000 | 1.000 |
| DDOS-HOIC | 0.547 | 0.788 | 0.998 | 1.000 | 1.000 | 1.000 | 1.000 | 1.000 |
| DDOS-LOIC-HTTP | 0.000 | 0.000 | 0.998 | 1.000 | 1.000 | 1.000 | 1.000 | 1.000 |
| DDOS-LOIC-UDP | 0.000 | 0.000 | 0.423 | 0.715 | 0.746 | 0.759 | 0.761 | 0.761 |
| DoS-GoldenEye | 0.000 | 0.000 | 0.828 | 0.995 | 0.996 | 0.999 | 1.000 | 1.000 |
| DoS-Hulk | 0.763 | 0.372 | 1.000 | 0.999 | 1.000 | 1.000 | 1.000 | 1.000 |
| DoS-SlowHTTPTest | 0.000 | 0.000 | 0.500 | 0.440 | 0.544 | 0.543 | 0.543 | 0.542 |
| DoS-Slowloris | 0.000 | 0.000 | 0.020 | 0.907 | 0.905 | 0.993 | 0.994 | 0.996 |
| FTP-BruteForce | 0.498 | 0.853 | 0.498 | 0.866 | 0.875 | 0.875 | 0.875 | 0.875 |
| SSH-Bruteforce | 0.000 | 0.000 | 0.999 | 1.000 | 1.000 | 1.000 | 1.000 | 1.000 |
| macro avg | 0.236 | 0.264 | 0.724 | 0.902 | 0.915 | 0.924 | **0.925** | **0.925** |
| weighted avg | 0.556 | 0.674 | 0.810 | 0.982 | 0.984 | 0.985 | **0.985** | **0.985** |

TABLE III
CIDS-NET RESULTS UNDER SCVIC-CIDS-2021 USING VARIOUS
NETWORK ENCODERS

|  | Baseline(XGB) | CIDS-Net using Different Network F | | | |
|---|---|---|---|---|---|
|  |  | Identity | FCN | FCN (CLoss) | TabNet |
| Benign | 0.9995 | 0.9995 | 0.9993 | 0.9963 | 0.9998 |
| Bot | 1.0000 | 0.9988 | 0.9978 | 0.9901 | 0.9994 |
| DDOS-HOIC | 1.0000 | 1.0000 | 0.9996 | 0.9998 | 1.0000 |
| DDOS-LOIC-HTTP | 0.9997 | 0.9997 | 0.9982 | 0.9989 | 0.9999 |
| **DDOS-LOIC-UDP** | **0.7609** | **0.9771** | **0.9884** | **0.9899** | **0.9927** |
| DoS-GoldenEye | 1.0000 | 0.9923 | 0.9914 | 0.9940 | 0.9996 |
| DoS-Hulk | 1.0000 | 0.9992 | 0.9998 | 0.9665 | 0.9988 |
| **DoS-SlowHTTPTest** | **0.5425** | **0.9228** | **0.9964** | **0.9233** | **0.9990** |
| DoS-Slowloris | 0.9936 | 0.9894 | 0.9929 | 0.9716 | 0.9993 |
| **FTP-BruteForce** | **0.8749** | **0.9721** | **0.9991** | **0.9713** | **0.9996** |
| SSH-Bruteforce | 0.9999 | 0.9998 | 0.9984 | 0.9994 | 0.9998 |
| **Macro Avg** | **0.9246** | **0.9864** | **0.9965** | **0.9819** | **0.9989** |
| **Weighted Avg** | **0.9848** | **0.9967** | **0.9990** | **0.9934** | **0.9998** |

encoder to a matrix $R^{100*768}$. The training and test sets are randomly split, and the number of samples are listed in Table I.

### B. CIDS-Net Results

This section discusses the outcomes of the CIDS-Nets using various network encoders. To establish a reasonable baseline, we begin by listing the outcomes of several machine/deep learning methods when only network-based features are employed in Table II. Due to the fact that network-based features dominate the final results, multiple network feature encoders are evaluated, including an Identity Layer (which means that the aggregator will take the network-based features as is), FCN, and TabNet. Additionally, this section evaluates the effect of C Loss and trivial loss.

Table II reports the baseline performance of various machine learning algorithms using only network-based features. Logistic Regression (LR), AdaBoost (AB) and Naive Bayes (NB) perform poorly (less than 90% macro F1 score) and exhibits bias toward majority classes.

Although Decision Tree performs slightly lower than XG-Boost and Random Forest on network-based features, as a simple, explainable, and fast machine learning model, it is employed for the following analytical purposes. Multiple ensemble methods are also used, among which, XGBoost and Random Forest produce nearly identical results and outperform all other approaches under study; thus, XGBoost's output serves as a representative baseline for comparing CIDS results.

Due to the fact that Fully Connected Networks can hardly outperform traditional machine learning techniques, this work also evaluates the TabNet, a deep learning model optimized for tabular datasets. However, it does not outperform XGBoost or Decision Tree.

Table III reports the performance of the CIDS-Net when various network encoders are used in conjunction with FCN as the aggregator. The macro F1 score is 0.986 if the network features are not encoded into latent states before being delivered to the FCN aggregator. As host-based features were applied, the macro F1 score increased by more than 5% in comparison to one of the best baselines (XGBoost). Three classes that experience the highest improvement levels are 'DDOS-LOIC-UDP', 'DoS-SlowHTTPPTest', and 'FTP-BruteForce'. The F1 scores for these classes are improved by 21.62%, 38.03%, and 9.72%, respectively. Even though these attacks have the same network patterns, they trigger different system responses such as failed FTP login and the number of HTTP heads. When the FCN is used as a network encoder, the macro F1 score increases by more than 1%, showing that the encoded latent states can function better with the FCN aggregator as additional parameters are provided. To investigate the effect of network size, we examine the impact of different number of FCN layers and neurons on macro average F1 score, when the CIDS-Net uses FCN as the network encoder and aggregator.

This work also examines the C loss. The main objectives of C Loss are to avoid training an additional NIDS and to incorporate a more complex network encoder, such as TabNet. Later in this section, this paper discusses the performance of employing a network encoder as a classifier. When C Loss is used, the CIDS performance is slightly diminished, but when TabNet (which also benefits from C Loss) is used as the network encoder, macro average F1 score is increased to 99.89% which is slightly (i.e., 0.2%) higher than the case when FCN is used as the network encoder.

Due to the possibility that the host-based features may not be available, the performance of network encoders is evaluated. When the FCN is employed as an aggregator, the FCN network encoder (without C Loss) performs poorly as a classifier, but after applying C Loss, the performance of the FCN network encoder increases from 7.46% to 76.56%. When TabNet (which also benefits from C Loss) is used as a network encoder, its performance is essentially identical to that when TabNet is trained exclusively with network-based features, demonstrating C Loss' usefulness. Nonetheless, even with C Loss and TabNet, the network encoder's performance does not outperform that of the XGBoost baseline. As a result, machine learning-based NIDS remains the preferred method, unless a deep learning model surpasses them on tabular datasets.

TABLE IV
CIDS-NET NETWORK ENCODER PERFORMANCE UNDER SCVIC-CIDS-2021

|  | XGBoost | CIDS-Net's Network Encoder | | |
|---|---|---|---|---|
|  |  | FCN | FCN (C Loss) | TabNet |
| Benign | 0.9995 | 0.0000 | 0.9668 | 0.9965 |
| Bot | 1.0000 | 0.0000 | 0.9885 | 0.9983 |
| DDOS-HOIC | 1.0000 | 0.0000 | 0.9897 | 0.9951 |
| DDOS-LOIC-HTTP | 0.9997 | 0.8202 | 0.9642 | 0.9997 |
| DDOS-LOIC-UDP | **0.7609** | 0.0000 | 0.3911 | 0.7365 |
| DoS-GoldenEye | 1.0000 | 0.0000 | 0.3649 | 0.9910 |
| DoS-Hulk | 1.0000 | 0.0000 | 0.9861 | 0.9991 |
| DoS-SlowHTTPTest | **0.5425** | 0.0000 | 0.4710 | 0.4978 |
| DoS-Slowloris | 0.9936 | 0.0000 | 0.6173 | 0.8830 |
| FTP-BruteForce | **0.8749** | 0.0000 | 0.7187 | 0.8493 |
| SSH-Bruteforce | 0.9999 | 0.0000 | 0.9628 | 0.9998 |
| macro avg | **0.9246** | 0.0746 | 0.7656 | 0.9042 |
| weighted avg | **0.9848** | 0.0517 | 0.9487 | 0.9796 |

## V. CONCLUSION

Two types of IDS (i.e., NIDS and HIDS) have been widely adopted and investigated while their combination has not been investigated enough. With this in mind, this paper has proposed a novel Combined IDS (CIDS) which can improve IDS performance by bringing together and aligning network and host-based data. A new dataset named SCVIC-CIDS-2021 has been constructed from the well-known benchmark dataset CIC-IDS-2018, and machine learning-based CIDS has been evaluated under SCVIC-CIDS-2021 in comparison to NIDS with network-based features as baselines. The highest performing baseline models have been shown as XGBoost and Random Forest, which achieve 92.5% in terms of macro F1 score. This work has also examined several combinations of network encoders (i.e., Identity, FCN, and TabNet) and loss functions (i.e., cross entropy loss and C Loss) for the CIDS-Net. When FCN is utilized as the aggregator, the CIDS-Net has been shown to perform optimally when TabNet is used as the network encoder, achieving a macro F1 score of 99.89%. The CIDS results have demonstrated that host-based features can significantly improve IDS performance.

Future work can integrate machine learning models into CIDS-Net to improve the performance on imbalanced classes. Moreover, the hyperparameters can be further tuned to balance the model complexity and performance.

## ACKNOWLEDGMENT

This work is supported in part by the Ontario Centre for Innovation (OCI) under ENCQOR 5G Project #31993.